\documentclass[letterpaper,english,aps,prb,amsmath,amssymb,reprint,author-year,author-numerical,floatfix,raggedfooter,superscriptaddress]{revtex4-1}

\usepackage[latin9]{inputenc}
\usepackage{mathtools}
\usepackage{bm}
\usepackage{amsmath}
\usepackage{amssymb}
\usepackage{graphicx}

\usepackage[caption=false]{subfig}

\usepackage{verbatim}

\makeatletter

\providecommand{\tabularnewline}{\\}

\usepackage{dcolumn}
\usepackage{bm}
\usepackage{booktabs}

\usepackage{xcolor}
\usepackage{soul}

\makeatother

\usepackage{babel}
\begin{document}

\title{Limitations of $\it{ab~initio}$ methods to predict the electronic-transport properties of two-dimensional materials: The computational example of 2H-phase transition metal dichalcogenides}

\author{Gautam~Gaddemane}

\affiliation{Department of Materials Science and Engineering, The University of
Texas at Dallas~\\
 800 W. Campbell Rd., Richardson, TX 75080, USA}

\author{Sanjay~Gopalan}

\affiliation{Department of Materials Science and Engineering, The University of
Texas at Dallas~\\
 800 W. Campbell Rd., Richardson, TX 75080, USA}
 
\author{Maarten~L.~{Van~de~Put}}

\affiliation{Department of Materials Science and Engineering, The University of
Texas at Dallas~\\
 800 W. Campbell Rd., Richardson, TX 75080, USA}
 
\author{Massimo~V.~Fischetti}

\affiliation{Department of Materials Science and Engineering, The University of
Texas at Dallas~\\
 800 W. Campbell Rd., Richardson, TX 75080, USA}
\email{max.fischetti@utdallas.edu.}

\date{\today}

\begin{abstract}
Over the last few years, $ab~initio$ methods have become an increasingly popular tool to evaluate intrinsic carrier transport properties in 2D materials. The lack of experimental information, and the progress made in the development of DFT tools to evaluate electronic band structures, phonon dispersions, and electron-phonon scattering matrix-elements, have made them a favored choice. However, a large discrepancy is observed in the literature among the $ab~initio$ calculated carrier mobility in 2D materials. Some of the discrepancies are a result of the physical approximations made in calculating the electron-phonon coupling constants and the carrier mobility. These approximations can be avoided by using a sophisticated transport model. However, despite using appropriate transport models, the uncertainty in the reported carrier mobility is still quite large in some materials. The major differences observed between these refined model calculations are the `flavors' of DFT (exchange-correlation functional, pseudopotential, and the effect of spin-orbit coupling) used. Here, considering several monolayer 2H-TMDs as examples, we calculate the low- and high-field transport properties using different `flavors' of DFT, and calculate a range for the electron mobility values. We observe that in some materials the values differ by orders of magnitude (For example, in monolayer WS$_{2}$ the electron low-field mobility varies between 37 cm$^{2}$/(V$\cdot$s) and 767 cm$^{2}$/(V$\cdot$s)). We analyze critically these discrepancies, and try to understand the limitations of the current $ab~initio$ methods in calculating carrier transport properties.
\end{abstract}

\keywords{Two-dimensional materials, electron-phonon interactions, deformation
potentials, Density Functional Theory}
\maketitle

\section{Introduction}

Two-dimensional (2D) materials have gained wide interest in the field of electronics as a potential channel material in field-effect transistors (FETs). The ability of these materials to confine carriers to atomically thin layers provides excellent electrostatic control and reduced short-channel effects. Moreover, their layered nature also reduces/eliminates deviations from ideality, such as surface roughness, dangling bonds, and interface states. Graphene\cite{Geim_2007, Bolotin_2008,Geim_2008}, silicene\cite{Vogt_2012,Roome_2014,Tao_2015,Li_2013}, , silicane\cite{Houssa_2011,Restrepo_2014,Low_2014,Khatami_2019}, germanene\cite{Houssa_2011,Roome_2014,Davila_2014}, phosphorene\cite{Gomez_2014,Li_2014,Liu_2014,Cao_2015,Doganov_2015,Gillgren_2014}, and monolayer transition metal dichalcogenides (TMD)\cite{Mak_2010,mos2_fet,Larentis_2012} are some of the most widely studied 2D materials. 

There has been a rapid rise in the development of technology to grow 2D materials\cite{lee2010wafer, kang2015high}, and fabricate FETs using them as channel material\cite{Tao_2015,Li_2014,Cao_2015,mos2_fet, Larentis_2012}. However, the experimental research is still in the early stages, and little is known about the intrinsic carrier transport properties. In fact, most of the predictions made are based on theoretical studies. 
\begin{figure}[tb]
\centering
\includegraphics[width=3.2in]{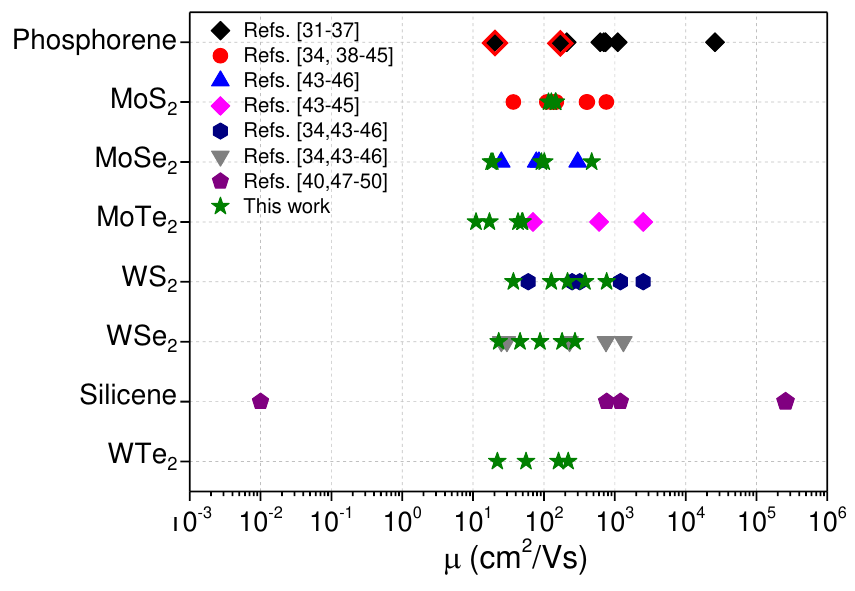}

\caption{Spread of the reported mobilities in literature for various 2D materials. Note that the x-axis has a logarithmic scale.}
\label{fig:mobility}
\end{figure}

The lack of experimental information, and timely progress made in $\it{ab~initio}$ (``first principles'') methods, especially in density functional theory (DFT), to calculate electronic band structures\cite{VASP4, Giannozzi_2009}, phonon dispersions\cite{Baroni_2001,phonopy}, and electron-phonon interactions\cite{giustino2017electron,Vandenberghe_2015,epw}, have made these methods a popular tool to calculate intrinsic carrier transport properties in 2D materials. However, when comparing the calculated intrinsic electron mobility values reported in the literature, a large discrepancy is observed among the results obtained via DFT-based calculations. Figure~\ref{fig:mobility} shows the spread of the values of the electron mobility reported in the literature calculated using DFT. The values range from 20 to 26000 cm$^{2}$/(V$\cdot$s) for phosphorene\cite{gaddemane_2018a, liao2015,jin2016,sohier2018,rudenko2016,trushkov2017,qiao2014}, 37 to 760 cm$^{2}$/(V$\cdot$s) for MoS$_{2}$\cite{gunst2016,li2015,li2013,kaasbjerg2012,sohier2018,zhao2018,rawat2018,zhang2014,huang2016}, 25 to 269 cm$^{2}$/(V$\cdot$s) for MoSe$_{2}$\cite{rawat2018,kim_tmd,zhang2014,huang2016}, 40 to 2526 cm$^{2}$/(V$\cdot$s) for MoTe$_{2}$\cite{rawat2018,zhang2014,huang2016}, 60 to 1103 cm$^{2}$/(V$\cdot$s) for WS$_{2}$\cite{sohier2018,kim_tmd,zhang2014,huang2016,rawat2018}, 25 to 705 cm$^{2}$/(V$\cdot$s) for WSe$_{2}$\cite{sohier2018,kim_tmd,rawat2018,zhang2014,huang2016}, and 0.01 to 250,000 cm$^{2}$/(V$\cdot$s) for silicene and germanene\cite{fischetti2016mermin,li2013,gaddemane2018b,shao2013,ye2014}.

These discrepancies can be explained, to some extent, by the physical models used for the transport calculations. It was previously shown in the case of phosphorene that using the Bardeen-Shockley deformation potentials\cite{bardeen1950} (constant deformation potentials) for scattering (thus ignoring the anisotropy of the electron-phonon matrix elements), and the semi-empirical Takagi formula\cite{takagi1994} for the mobility calculations (thus ignoring optical and intervalley scattering), leads to serious errors and overestimation of the carrier mobility.\cite{gaddemane_2018a} By considering full electron-phonon matrix-elements (dependent on the electron wavevector $\bf{k}$ and the phonon wavevector $\bf{q}$), and full electron and phonon dispersions (full-bands), the electron mobility in phosphorene was found to be on the lower end of the range shown (highlighted) in Fig.~\ref{fig:mobility} (21  cm$^{2}$/(V$\cdot$s)\cite{gaddemane_2018a} and 170 cm$^{2}$/(V$\cdot$s)\cite{liao2015}). In free-standing 2D materials like silicene and germanene, which lack the horizontal mirror ($\sigma_{\rm h}$) symmetry, it was shown that the coupling of carriers with the out-of-plane acoustic phonon modes  (ZA) is extremely strong, an effect that leads to a significantly low carrier mobility\cite{fischetti2016mermin}. The calculations which predict mobility values in the higher end of the range in Fig.~\ref{fig:mobility} for these materials were performed by ignoring the scattering with ZA phonons, and using constant deformation potentials and Takagi formula for the scattering and mobility calculations, respectively.

 Nevertheless, even when comparing calculations which were performed using appropriate physical models, a significant discrepancy in the carrier mobility  is still observed. The major differences between these DFT-based calculations were the pseudopotentials and the exchange-correlation functionals used. For example, in the case of phosphorene, the authors of both Refs.~\onlinecite{gaddemane_2018a} and~\onlinecite{liao2015} performed their calculations using full electron-phonon matrix-elements, and full electron and phonon dispersions, but a different exchange-correlation functional and pseudopotentials. This resulted in an order of magnitude difference in their electron mobility.

Monolayer transition metal dichalcogenides (2H-TMD) with 2H- phase, such as MoS$_{2}$, MoSe$_{2}$, MoTe$_{2}$, WS$_{2}$, WSe$_{2}$, and WTe$_{2}$,  are by far the most widely studied 2D materials due to their layered nature and wide band gap. As shown in Fig.~\ref{fig:mobility}, there is a large variation in the reported mobility values for these materials as well. Even restricting our attention only to the calculations which were performed using appropriate numerical models (including full electron and phonon dispersion, full electron-phonon matrix-elements, acoustic and optical phonon scattering, and intra- and inter-valley scattering processes), a significant difference in the calculated electron mobility is observed. Similar to what we saw in phosphorene, the differences arise from the different `flavors' (pseudopotentials, exchange-correlation functional, spin-orbit effects)  of DFT used. 

In this paper, we present a detailed DFT based transport study on monolayer 2H-TMDs (MoS$_{2}$, MoSe$_{2}$, MoTe$_{2}$, WSe$_{2}$, and WTe$_{2}$). We start by obtaining the relaxed crystal structure, followed by the calculation of the electronic band structure, the phonon dispersion, and the electron-phonon matrix-elements using different exchange-correlation functionals, pseudopotentials, and with and without spin-orbit coupling (SOC). We then use this information as an input to a full-band Monte Carlo program that calculates low- and high-field transport properties. 

In Sec.~\ref{sec:Methods} we present a brief overview of the computational methods used to study the low- and high-field electron transport from DFT based methods. In Sec.~\ref{sec:results}, we present the results for the 2H-TMDs mentioned above, focusing in particular to WS$_{2}$.

\section{Computational method}
\label{sec:Methods} 

\begin{table}
\centering
\caption{Computational parameters used in the DFT calculations. $\it {E}$$_\mathrm{k}$ is the kinetic energy and `SCF' stands for `self-consistent field'.}
\label{tab:Comput_dft}
\begin{ruledtabular}
\begin{tabular}{c|c}
Parameters  & Quantum ESPRESSO \tabularnewline
\hline 
$\it {E}$$_\mathrm{k}$ cutoff &60~Ry\tabularnewline

Ionic minimization threshold  & $10^{-6}$~Ry\tabularnewline

SCF threshold & $10^{-12}$~Ry\tabularnewline

$\bold{k}$-points mesh&$12\times12\times1$ \tabularnewline

\end{tabular}
\end{ruledtabular}



\end{table}

In this section, we give a brief explanation on the methods used, and provide the necessary computation details. For a more detailed explanation, we direct the readers to our previous work presented in Ref.~\onlinecite{gaddemane_2018a}.

The electronic band structures for 2H-TMDs are obtained using the density functional theory (DFT) formalism as implemented in the Quantum ESPRESSO (QE)\cite{Giannozzi_2009} package with the local density approximation (LDA)\cite{LDA} and the Perdew-Burke-Enzerhoff  generalized-gradient approximation (GGA-PBE)\cite{PBE} for the exchange-correlation functional, and the norm-conserving Vanderbilt (ONCV)\cite{oncv} pseudopotentials and standard solid state pseudopotentials (SSSP)\cite{sssp} for the pseudopotentials of each constituent element. We have performed our calculations with and without SOC. The atomic structure for these materials is found by minimizing the total energy with respect to the lattice constants and ionic positions. The computational parameters used in these calculations are shown in Table~\ref{tab:Comput_dft}. The band structure is calculated and tabulated on a fine mesh, 201$\times$201$\times$1, covering a rectangular section that inscribes the triangular irreducible wedge of the hexagonal first Brillouin zone.

For the calculation of the phonon dispersion and of the electron-phonon matrix elements, we have also used QE\cite{Baroni_2001,giustino2017electron}, augmented by the Electron-phonon Wannier (EPW)\cite{epw} software package which uses the density functional perturbation theory (DFPT) formalism. These quantities are initially calculated on a coarse $\bf k$ (12$\times$12$\times$1) and $\bf q$ (6$\times$6$\times$1) mesh and interpolated on a fine $\bf k$ (30$\times$30$\times$1) and $\bf q$ (30$\times$30$\times$1) mesh using maximally-localized Wannier-functions. Both coarse and fine meshes span the entire Brillouin zone. The phonon dispersion and electron-phonon matrix elements on the fine mesh are finally interpolated on the band structure mesh using a bilinear interpolation. The electron-phonon scattering rates are calculated using Fermi's golden rule, and tabulated on the same mesh of $\bf k$-points used to tabulate the band structure. 

Using this information, we solve the Boltzmann transport equation employing full-band Monte Carlo simulations. Scattering with ZA phonons is ignored in these calculations, since 2H-TMDs have horizontal mirror ($\sigma_{\rm h}$) symmetry, so the process is forbidden to the first-order. The electron mobility, $\mu_{\theta}$ along the direction $\theta$ is extracted from the diffusion constant $D_{\theta}$, rather than from the velocity-field characteristics since the former is less affected by stochastic noise\cite{jacoboni1983}.

We repeat the whole procedure of relaxing the crystal structure, calculating the band structure, the phonon dipsersion, and the electron-phonon matrix elements, and performing the Monte Carlo simulation for each different `flavor' of DFT.

\section{Results and Discussion}
\label{sec:results}

In this section, we study the dependence of the transport properties on the pseudopotential (Sec.~\ref{ss:pseudo_comp}), the inclusion of spin-orbit-coupling (Sec.~\ref{ss:soc_comp}) and the exchange and correlation functional (Sec.~\ref{ss:xc_comp}) for WS$_{2}$, as an example. In Sec.~\ref{ss:other_mtl}, we discuss the results for all studied 2H-TMDs.

\subsection{Psuedopotential comparison study }
\label{ss:pseudo_comp}

\begin{figure*}
\centering
\includegraphics[]{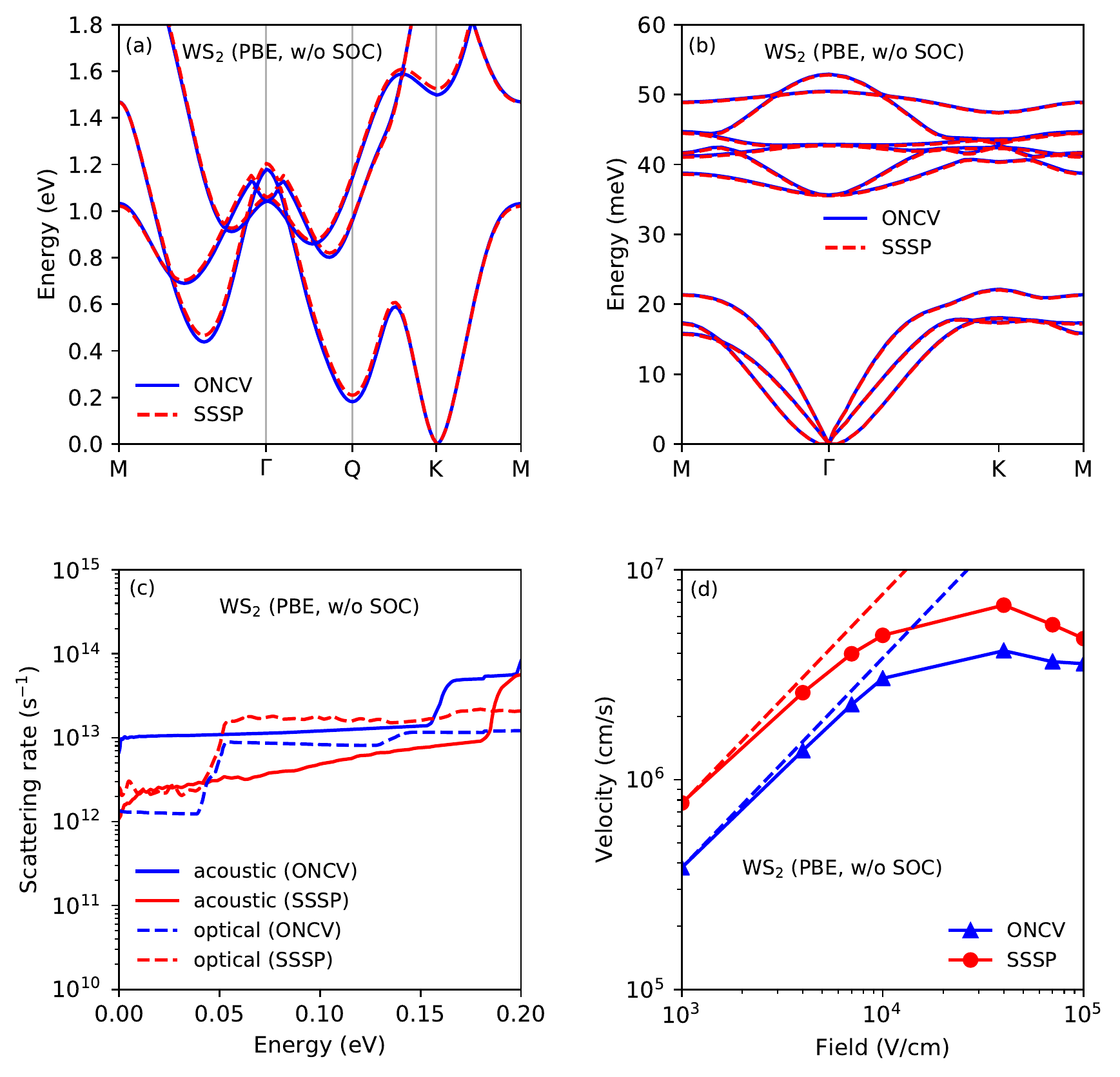}

\caption{(a) The electronic bandstructure, (b) phonon dispersion, (c) electron-phonon scattering rates, and (d) velocity-field characteristics for WS$_{2}$, calculated using the ONCV pseudopotentials and the SSSP for the psudopotentials, the GGA-PBE exchange-correlation functional, and without spin-orbit coupling.}
\label{fig:pseudo}
\end{figure*}

First, we compare the intrinsic electron transport properties of WS$_{2}$, calculated using two different pseudopotentials: the ONCV pseudopotentials and the SSSP for the DFT and DFPT calculations. We selected the ONCV pseudopotentials and the SSSP, since they are well tested, mature and often used in the QE community. Both sets of calculations were performed using the GGA-PBE exchange-correlation functional and by ignoring the SOC effect (for now).

Figure~\ref{fig:pseudo}a and~\ref{fig:pseudo}b shows the electronic band structure and the phonon dispersion for WS$_{2}$, calculated using both pseudopotentials, and plotted along high-symmetry directions. Regardless of the choice of pseudopotential, the conduction band minima is located at the three-fold degenerate K valley with an isotropic effective mass of  0.30. A six-fold degenerate satellite valley called the Q valley is located along the K--$\Gamma$ direction with a longitudinal effective mass of 0.60, and a transverse effective mass of 0.77, when calculated using both the ONCV pseudopotentials and SSSP. The major difference between the conduction bands is the energy difference between the K and Q valleys, $\Delta {\rm E_{\rm KQ}}$ which is approximately 182 and 210 meV when calculated using the ONCV pseudopotentials and the SSSP, respectively. The phonon dispersion is identical in both the cases. 

However, when we compare the scattering rates (Fig.~\ref{fig:pseudo}c), the acoustic phonons scattering rate calculated using the ONCV pseudopotentials is approximately one order of magnitude larger than the one calculated using the SSSP at low energies. Despite having a similar band structure and phonon dispersion, the electron-phonon matrix-elements calculated using different pseudopotentials differ by approximately an order of magnitude. Although the eigenvalues (electron and phonon dispersion) obtained are similar, the eigenvectors (wavefunctions and polarization vectors) obtained can vary significantly depending on the pseudopotential considered, which results in different scattering matrix-elements. The step-like increase observed at higher energy corresponds to the onset of  K--Q intervalley scattering. Since the energy difference between the minima of the K and Q valleys, $\Delta {\rm E_{\rm KQ}}$ is lower when calculated using the ONCV pseudopotentials, we observe the onset of intervalley scattering at a lower energy compared to the one calculated using the SSSP. However, for optical phonons, the calculation performed using the SSSP results in a slightly larger scattering rate compared to the one obtained using the ONCV pseudopotentials. Intervalley scattering is also observed for optical phonons but it is much weaker compared to what we observed for acoustic phonons, regardless of the pseudopotentials used.

The room-temperature electron mobility, obtained from Monte Carlo simulations, is 380 and 767 cm$^{2}$/(V$\cdot$s) when calculated using the ONCV pseudopotentials and the SSSP, respectively. The significantly lower mobility obtained using the ONCV pseudopotentials is due to a larger intravalley acoustic phonons scattering, and the early onset of  K--Q intervalley scattering, compared to using the SSSP.  While the dispersion relation of both electrons and holes are largely independent of the chosen pseudopotential (with the notable exception of the sattelite valley energy), the predicted mobility differs by a factor of two.

The velocity-field characteristics for WS$_{2}$, obtained using both the ONCV pseudopotentials and the SSSP, for a field applied along the zigzag, are shown in Fig.~\ref{fig:pseudo}d. The zigzag direction corresponds to the direction along $\Gamma$ to K in $\bf k$-space. The mobility calculated from the linear region of the curves is in agreement with the value obtained from the diffusion constant. We observe  negative differential mobility for both the ONCV pseudopotential and the SSSP cases. This is due to significant transfer of electrons to the Q valley (larger effective mass thus lower velocity) at high fields. However, the peak velocity reached is approximately 4$\times$10$^{6}$ cm/s and 6$\times$10$^{6}$ cm/s for ONCV pseudoptentials and SSSP, respectively.

\subsection{Spin-orbit coupling}
\label{ss:soc_comp}

\begin{figure*}
\centering
\includegraphics[]{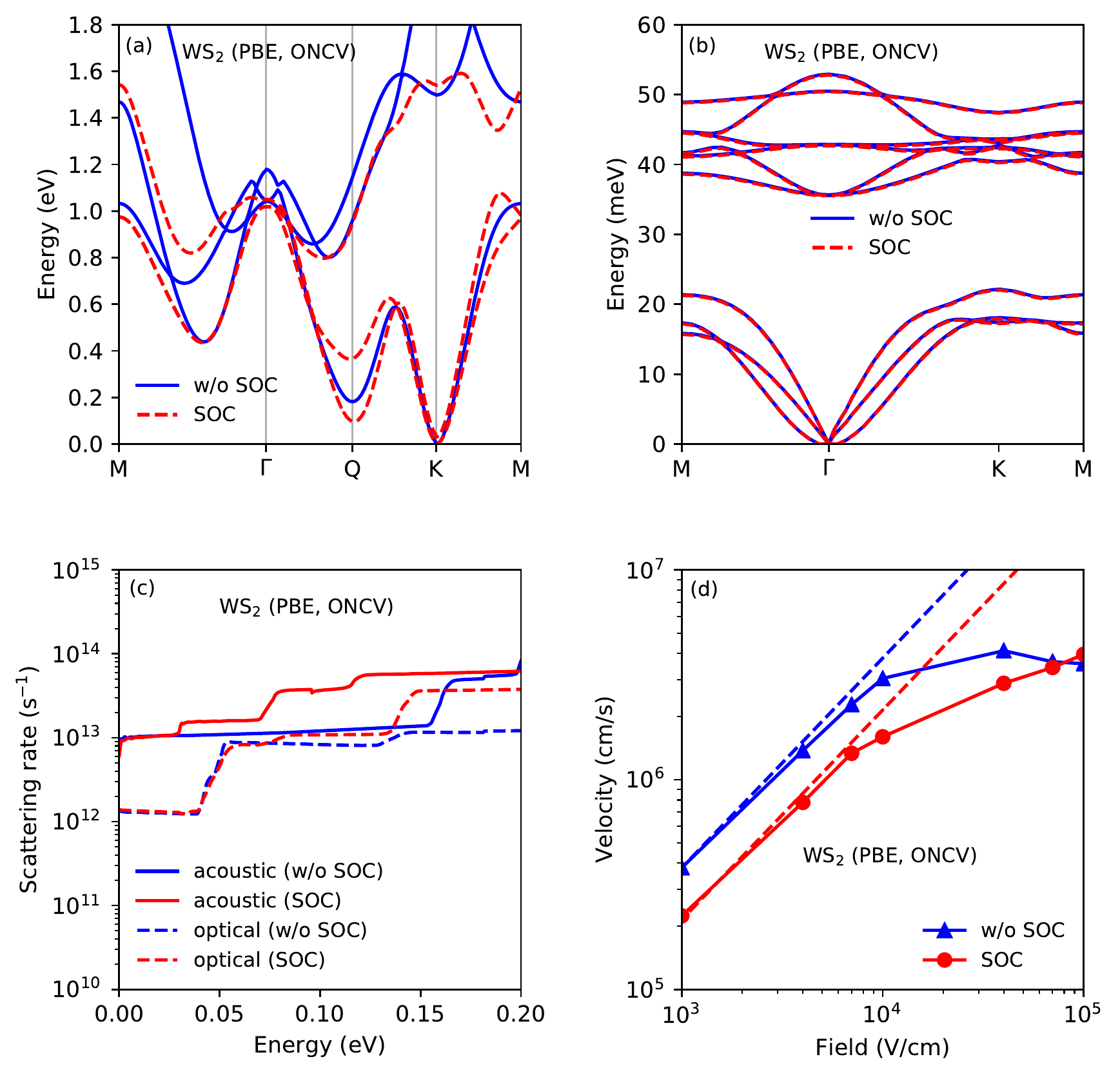}

\caption{(a) The electronic bandstructure, (b) phonon dispersion, (c) electron-phonon scattering rates, and (d) velocity-field characteristics for WS$_{2}$, calculated using the ONCV pseudopotentials, the GGA-PBE exchange-correlation functional, with and without spin-orbit coupling.}
\label{fig:soc}
\end{figure*}

Next, we compare the intrinsic transport properties of WS$_{2}$ calculated with and without the spin-orbit coupling (SOC) effect. The calculations were performed using the ONCV pseudopotentials and GGA-PBE exchange-correlation functional. The ONCV pseudopotentials were chosen because they offer a ``relativistic" version, optimized for use with the SOC-model in QE.

Figure~\ref{fig:soc}a and~\ref{fig:soc}b show the electronic band structure and the phonon dispersion for WS$_{2}$ calculated with and without SOC, and plotted along the symmetry directions. In the presence of SOC, the conduction band splitting is 30 and 250 meV at the K and Q symmetry points, respectively. The valley separation $\Delta {\rm E_{\rm KQ}}$ reduces to 97 meV from 182 meV in the presence of SOC, whereas the effective mass increases to 0.35 from 0.30 along the  K valley, and decreases to 0.54 from 0.60 along the logitudinal direction of the Q valley, when SOC is included in the calculations. However, as expected for materials without global spin-polarization, the phonon dispersion is left unaffected.

When comparing the scattering rates (Fig.~\ref{fig:soc}c), acoustic phonons dominate the scattering processes in both the cases at low energies. However, the onset of intervalley K--Q scattering occurs at a lower energy when SOC is included because of a lower $\Delta {\rm E_{\rm KQ}}$ value. The optical phonons scattering rates are similar in both cases at low energies but, similar to acoustic phonons, the step-like increase due to intervalley scattering occurs at different energies.

The room-temperature electron mobility obtained from Monte Carlo simulations reduces to 215 cm$^{2}$/(V$\cdot$s) from 380 cm$^{2}$/(V$\cdot$s) in the presence of SOC. The reduction in mobility is due to the early onset of K--Q intervalley scattering.

Finally, we show in Fig.~\ref{fig:soc}d the velocity-field characteristics for WS$_{2}$ for a field applied along the zigzag direction, comparing the calculations performed with and without SOC. Negative differential mobility is observed for the calculation performed without SOC with a peak velocity of approximately 4$\times$10$^{6}$ cm/s. However, when calculated with the SOC effect, due to its lower mobility, we do not observe velocity saturation or negative differential mobility even at high-fields of 10$^{5}$ V/cm.

\subsection{Exchange-correlation functional}
\label{ss:xc_comp}
\begin{figure*}
\centering
\includegraphics[]{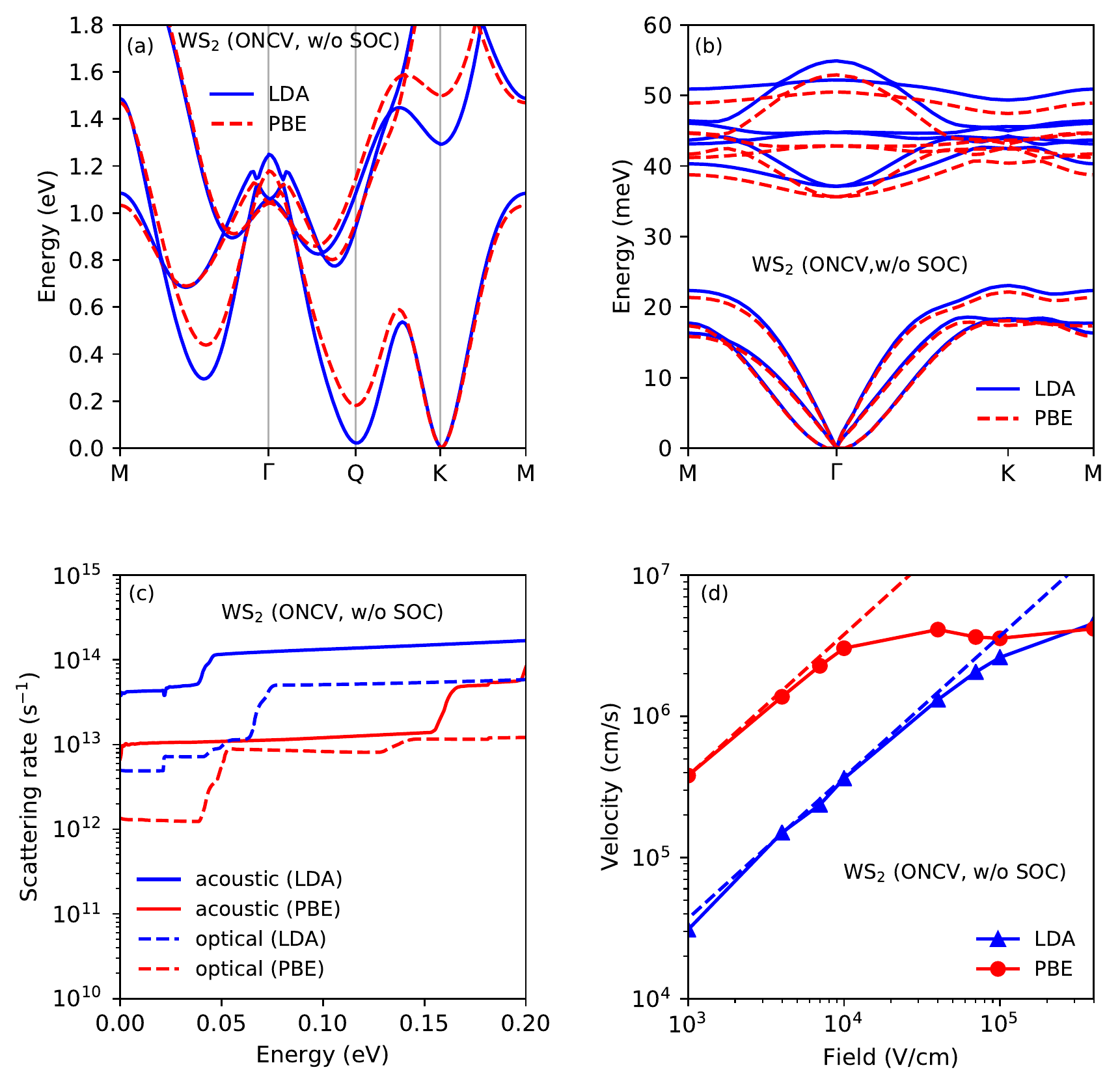}

\caption{(a) The electronic bandstructure, (b) phonon dispersion, (c) electron-phonon scattering rates, and (d) velocity-field characteristics for WS$_{2}$, calculated using the ONCV pseudopotentials, the LDA and the GGA-PBE exchange-correlation functional, and without spin-orbit coupling.}
\label{fig:xc_nsoc}
\end{figure*}

As a final variation, we compare the intrinsic electron transport properties of WS$_{2}$ calculated by using two different approximations to the exchange-correlation functional, the local density approximation (LDA) and the generalized-gradient approximation as proposed by Perdew, Burke and Enzerhof (GGA-PBE). The calculations were performed using the ONCV pseudopotentials, both with and without the SOC effect

We first discuss the results for the calculation performed without SOC, which are shown in Figs.~\ref{fig:xc_nsoc}. We show, in Figs.~\ref{fig:xc_nsoc}a and~\ref{fig:xc_nsoc}b, the  electronic band structure and the phonon dispersion for WS$_{2}$, comparing the calculations performed using both the LDA and the GGA-PBE exchange-correlation functional. The major difference seen in the band structure is the energy splitting between the valleys, $\Delta {\rm E_{\rm KQ}}$, which is approximately 22 and 182 meV for the LDA and GGA-PBE cases, respectively. The isotropic effective mass obtained for the K valley is 0.30 in both the calculations. However, the longitudinal and transverse effective masses for the Q valley are 0.55 and 0.74 when using the LDA, and 0.60 and 0.77 when using the GGA-PBE, for the exchange-correlation functional. The acoustic phonons dispersion is identical near the $\Gamma$ symmetry point for both the cases. However, at shorter wavelength, we obtain  slightly lower energies when using the LDA compared to the GGA-PBE for the exchange-correlation functional. Optical phonons have lower energies for all range of $\bf q$ when using LDA compared to GGA-PBE.

When we compare the scattering rates (Fig.~\ref{fig:xc_nsoc}c), the acoustic phonons scattering rate obtained when using the LDA exchange-correlation functional is larger than the one obtained when using the GGA-PBE exchange-correlation functional. Since $\Delta {\rm E_{\rm KQ}}$ is lower for the LDA case compared to the GGA-PBE case, we observe the onset of intervalley scattering at a lower energy for the former compared to the latter. Even for optical phonons, the scattering rates are higher when using LDA compared to using GGA-PBE.

The room-temperature electron mobility is 37 and 380 cm$^{2}$/(V$\cdot$s) when using the LDA and GGA-PBE, respectively. The significantly lower mobility obtained for the LDA case is due to larger intravalley acoustic phonons scattering and early onset of intervalley scattering compared to the GGA-PBE case.

 We show in Fig.~\ref{fig:xc_nsoc}d, the velocity-field characteristics for WS$_{2}$, for a field applied along the zigzag direction, when using the LDA and the GGA-PBE exchange-correlation functional. For the calculation performed using the GGA-PBE exchange-correlation functional, we observe negative differential mobility at high fields with a peak velocity of approximately 4$\times$10$^{6}$ cm/s. However, for calculations performed using LDA exchange-correlation functional, due to a lower mobility, we do not observe neither velocity saturation nor negative differential mobility even at high-fields of 10$^{5}$ V/cm. 

\begin{figure*}
\centering
\includegraphics[]{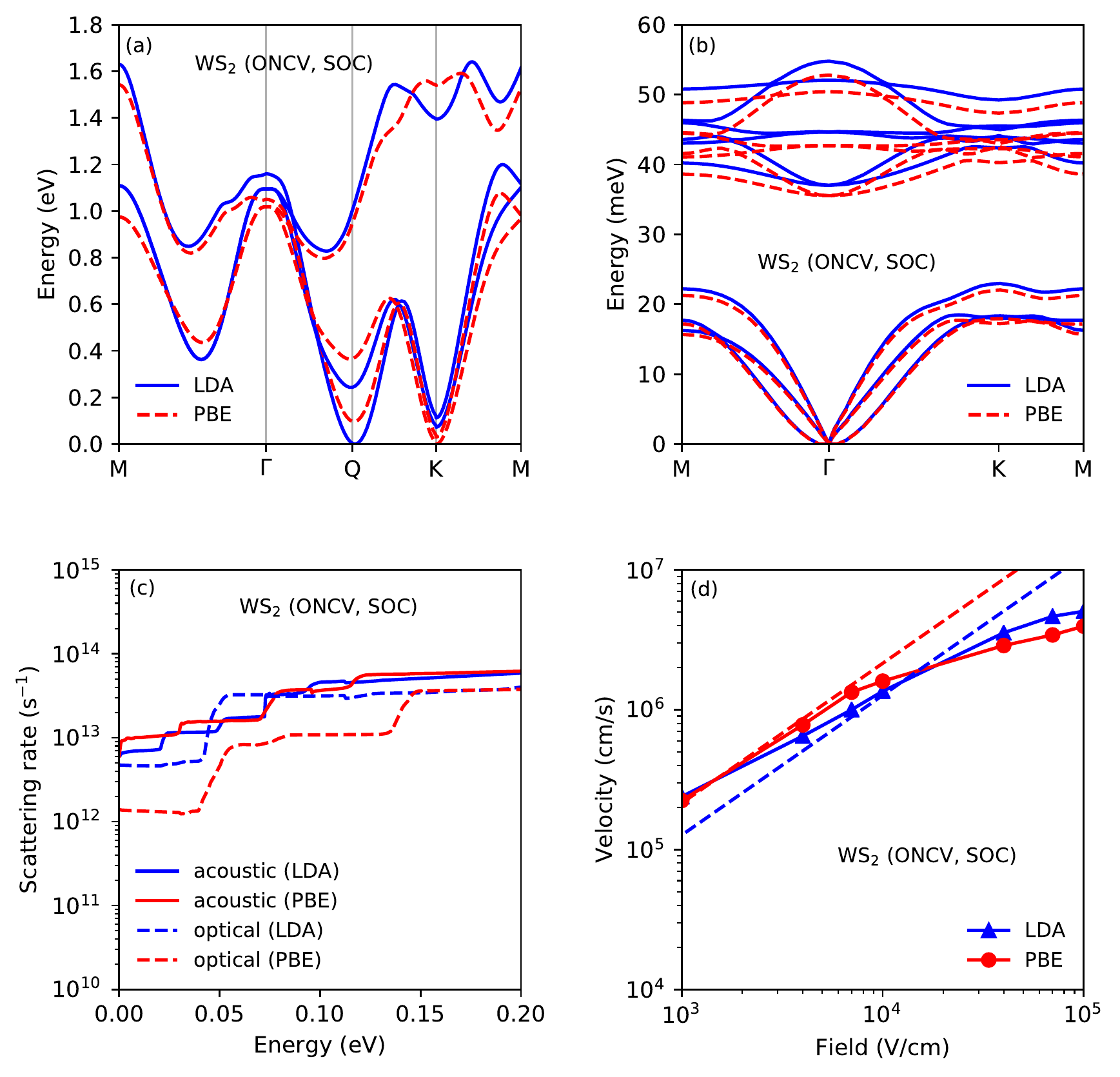}

\caption{(a) The electronic bandstructure, (b) phonon dispersion, (c) electron-phonon scattering rates, and (d) velocity-field characteristics for WS$_{2}$, calculated using the ONCV pseudopotentials,the LDA and the GGA-PBE exchange-correlation functional, and with spin-orbit coupling.}
\label{fig:xc_soc}
\end{figure*}

In Fig.~\ref{fig:xc_soc}, we show the results for a similar calculations but performed with the SOC effect. Due to the splitting of bands, we observe a change in the $\Delta {\rm E_{\rm KQ}}$ value compared to the values obtained for the calculations performed without the SOC effect. As seen in Fig.~\ref{fig:xc_soc}a, for the calculations performed using the GGA-PBE exchange-correlation functional, $\Delta {\rm E_{\rm KQ}}$ reduces to 97 meV from 182 meV with the conduction band minima still being at the K valley. However, when using the LDA for the exchange-correlation functional, the conduction band minima changes to the Q valley, with the Q valley being lower than the K valley by 73 meV. The effective mass obtained for the  K valley for both the LDA and GGA-PBE cases is not affected by the SOC effect. However, the effective mass decreases for the Q valley in both the LDA and the PBE cases (The effective masses along the longitudinal and transverse directions for the Q valley are 0.48 and 0.72 when using the LDA for the exchange-correlation functional, and 0.54 and 0.77 when using the PBE for the exchange-correlation functional). However, the phonon dispersion remains unchanged in the presence of SOC (Fig.~\ref{fig:xc_soc}b.     

Figure~\ref{fig:xc_soc}c shows the scattering rates, comparing the calculations performed using the LDA and the GGA-PBE for the exchange-correlation functional, in the presence of SOC. The trend in the results is similar to what we observed when the calculations were performed without SOC (The LDA case has a larger scattering rate compared to the GGA-PBE case). However, when using the LDA for the exchange-correlation functional, the acoustic phonon scattering rates are smaller when calculated with SOC compared to the scattering rates calculated without SOC, at low energies. This is due to the fact that the six-fold degenerate Q valley becomes the conduction band minimum with SOC activated, which has a smaller electron-phonon scattering matrix-elements compared to the three-fold degenerate K valley, leading to lower scattering rates. However, the onset of intervalley scattering between the K to Q valley occurs at a higher energy compared to the calculations performed without SOC  because of the higher $\Delta {\rm E_{\rm KQ}}$ value. The effect of SOC on the scattering rates, when using the GGA-PBE as the exchange correlation functional has already been discussed above in Sec.~\ref{ss:soc_comp}.

The room temperature electron mobility obtained is 127 and 215 cm$^{2}$/(V$\cdot$s) for the LDA and the GGA-PBE cases, respectively, when calculated in the presence of the SOC effect. The electron mobility increased for the calculation performed using the LDA for the exchange-correlation functional due to lower effective mass in the Q valley, and a later onset of the K and Q intervalley scattering. However, as discussed in Sec.~\ref{ss:soc_comp}, the electron mobility decreases for the calculation performed using the GGA-PBE as the exchange correlation functional in the presence of SOC due to early onset of intervalley scattering.

We show in Fig.~\ref{fig:xc_soc}d the velocity-field characteristics for WS$_{2}$ for a field applied along the zigzag direction, comparing the LDA and the GGA-PBE cases in the presence of SOC. We do not observe velocity saturation or negative differential mobility for both the LDA and the GGA-PBE cases, when calculated with SOC, even at high fields of 10$^{5}$ V/cm, in contrast to our calculations ignoring SOC.

\subsection{Intrinsic transport results for other 2H-TMDs}
\label{ss:other_mtl}

\begin{table}
\caption{Theoretical calculations of the electron mobility at 300 K,
$\mu_{{\rm e}}$, obtained from different `flavors' of DFT for all considered 2H-TMDs.}
\begin{ruledtabular}
\begin{tabular}{lccccc}
 $\mu_{{\rm e}}$  & SSSP-PBE & \multicolumn{2}{c}{ONCV-PBE } & \multicolumn{2}{c}{ONCV-LDA } \tabularnewline
 (cm$^{2}$/(V$\cdot$s)& w/o SOC  & w/o SOC  & w SOC  & w/o SOC  & w SOC \tabularnewline
\hline 
MoS$_{2}$  & 127  & 147  & 145 &127 &116  \tabularnewline
MoSe$_{2}$  & 78  & 92 & 101  & 18 & 19 \tabularnewline
MoTe$_{2}$  & 49  & 43  & 50  & 17   & 11\tabularnewline
WS$_{2}$  & 767  & 380 & 215  & 37 & 127  \tabularnewline
WSe$_{2}$  & 275  & 180  & 46 &  23& 88  \tabularnewline
WTe$_{2}$  & 220  & 220  & 161  & 56  & 22  \tabularnewline
\end{tabular}
\end{ruledtabular}

\label{tab:mu} 
\end{table}

\begin{table}
\caption{Energy separation between K and Q valleys obtained from different `flavors' of DFT calculated for all considered 2H-TMDs. * indicantes the calculated bandgap is indirect (conduction band minima at $\bf Q$).}
\begin{ruledtabular}
\begin{tabular}{lccccc}
$\Delta {\rm E_{\rm KQ}}$   & SSSP-PBE & \multicolumn{2}{c}{ONCV-PBE } & \multicolumn{2}{c}{ONCV-LDA } \tabularnewline
 (meV)& w/o SOC  & w/o SOC  & w SOC  & w/o SOC  & w SOC \tabularnewline
\hline 
MoS$_{2}$  & 270  & 265  & 231 &100 &71  \tabularnewline
MoSe$_{2}$  & 155  & 151 & 152  & 23$^{*}$ & 18$^{*}$ \tabularnewline
MoTe$_{2}$  & 155  & 170  & 177  & 12   & 26\tabularnewline
WS$_{2}$  & 210  & 182 & 97  & 22 & 73$^{*}$  \tabularnewline
WSe$_{2}$  & 124  & 117  & 41 &  74$^{*}$& 135$^{*}$  \tabularnewline
WTe$_{2}$  & 64  & 166  & 181  & 60  & 55  \tabularnewline
\end{tabular}
\end{ruledtabular}

\label{tab:Ekq} 
\end{table}

We studied a wide range of 2H-TMDs: MoS$_{2}$, MoSe$_{2}$, MoTe$_{2}$, WSe$_{2}$, and WTe$_{2}$, and   the electron mobility in these materials calculated for the various cases discussed above are shown in Table~\ref{tab:mu}. Table~\ref{tab:Ekq} lists the energy difference between the valleys, $\Delta {\rm E_{\rm KQ}}$ for these materials, calculated using different `flavors' of DFT used.
Similar discrepancies to those observed in WS$_{2}$ were also observed for these materials. In the Supplementary Material we present figures that illustrate in detail this state of affairs. We can summarize our results as follows: For the Mo-based TMDs, the effect of  SOC is seen to be negligible. However, in the heavier W-based materials, SOC has a significant effect on the energy difference between the valleys, which determines the onset of intervalley scattering and therefore has a large influence on the electron mobility. We also observe that the calculations performed using the LDA exchange-correlation functional give a lower $\Delta {\rm E_{\rm KQ}}$ leading to larger intervalley scattering, and hence lower electron mobility in all the materials considered.

 Until the availability of further experimental evidence on the accuracy of the band structure and the electron-phonon coupling constants, $ab~initio$ methods can at best provide a range for the carrier mobility.

\section{Conclusion}
The discrepancy found in the literature on the first principles calculations of electron mobility in 2D materials has motivated us to analyze critically the reasons for this confusion. In our previous work, we had shown that some of these calculations overestimate the mobility by improperly treating the electron-phonon interaction, and ignoring various scattering processes. However, even considering only the calculations performed using appropriate physical models, a large discrepancy still exists among the theoretically predicted electron mobility in 2D materials. This can be attributed to the `flavors' of  DFT used in the calculations. Here, considering 2H-TMDs as examples, we have shown a comparative study by calculating intrinsic carrier properties in these materials using different pseudopotentials and exchange-correlation functionals. We found that using different pseudopotentials and exchange-correlation functionals result in different band structures, phonon dispersion and the electron-phonon scattering matrix-elements. The calculation performed using the LDA for the exchange-correlation functional results in a lower electron mobility compared to the calculation performed using the GGA- PBE, due to a lower energy difference between the K and the Q valleys for the former case, resulting in a significantly larger contribution of intervalley scattering. We have also studied the effect of spin-orbit coupling, and we found that it becomes extremely important for W-based TMDs due to its larger atomic mass compared to the Mo-based TMDs.

Over the past decades, DFT has improved markedly, making $\it{ab~initio}$ calculations feasible in a large array of applications. However, as we have found in our current work, one should carefully consider the approximations and limitations of the theory and its implementations. It is well known that, as a ground-state theory, DFT's accuracy is limited when treating excited (conduction) states of the system, which are of the utmost importance to electronic transport. We have shown that transport properties are highly sensitive to variations of the electronic dispersion in the meV-range, as well as to the accuracy of the eigenvectors of the electron and phonon modes, properties that vary significantly among the different approximations made in DFT.

It is unclear, at this time, if these deficiencies can be resolved within the DFT framework, or even in more advanced models, such as the GW approximation. Predicting reliable results would require further experimental evidence, but due to the infancy of the technology for 2D materials, we can at best provide a range for the intrinsic mobility in 2D materials from $\it{ab~initio}$ methods. Unfortunately, in materials like WS$_{2}$, this range may vary upto two orders of magnitude.

\acknowledgments{This work has been supported in part by the Semiconductor Research Corporation (SRC nCORE) and Taiwan Semiconductor Manufacturing Company, Ltd (TSMC).}

\bibliographystyle{apsrev4-2}
\bibliography{tmd_bib}

\end{document}